\documentclass[final,5p,times,twocolumn]{elsarticle}

\usepackage{graphicx}

\usepackage{amssymb}
\usepackage{amsthm}

\usepackage{amsmath}

\usepackage{dcolumn}
\usepackage{bm}
\usepackage{here}  
\usepackage{color}

\usepackage{bm}
\usepackage{here}
\usepackage{color}

\definecolor{mygreen}{rgb}{0.2,0.8,0.2}

\begin{document}

\begin{frontmatter}


\title{Localization of 
disordered harmonic chain with long-range correlation
}




 \author{Hiroaki S. Yamada}
\address{Yamada Physics Research Laboratory, Aoyama 5-7-14-205, Niigata 950-2002, Japan}

\begin{abstract}
In the previous paper [Yamada, Chaos, Solitons $\&$ Fractals, {\bf 109},99(2018)], 
we investigated localization properties of one-dimensional disordered 
electronic system with long-range
correlation generated by modified Bernoulli (MB) map. 
In the present paper, we report localization properties of  phonon in 
disordered harmonic chains generated by the MB map. 
Here we show that  
Lyapunov exponent becomes positive definite for almost all frequencies $\omega$ 
except $\omega=0$, 
and the  $B-$dependence 
changes to exponential decrease for  $B > 2 $, 
where $B$ is a correlation parameter of the MB map.
The distribution of the Lyapunov exponent of the phonon amplitude has a
slow convergence,  different from that of  uncorrelated disordered
systems obeying a normal central-limit theorem. 
Moreover, we calculate the phonon dynamics 
in the MB chains. We show that the time-dependence 
of spread in the phonon amplitude and energy wave packet 
changes from that  in the disordered chain 
to that in the periodic one, as the correlation parameter $B$ increases.
\end{abstract}


\end{frontmatter}


\def\ni{\noindent}
\def\nn{\nonumber}
\def\bH{\begin{Huge}}
\def\eH{\end{Huge}}
\def\bL{\begin{Large}}
\def\eL{\end{Large}}
\def\bl{\begin{large}}
\def\el{\end{large}}
\def\beq{\begin{eqnarray}}
\def\eeq{\end{eqnarray}}

\def\eps{\epsilon}
\def\th{\theta}
\def\del{\delta}
\def\omg{\omega}

\def\e{{\rm e}}
\def\exp{{\rm exp}}
\def\arg{{\rm arg}}
\def\Im{{\rm Im}}
\def\Re{{\rm Re}}

\def\sup{\supset}
\def\sub{\subset}
\def\a{\cap}
\def\u{\cup}
\def\bks{\backslash}

\def\ovl{\overline}
\def\unl{\underline}

\def\rar{\rightarrow}
\def\Rar{\Rightarrow}
\def\lar{\leftarrow}
\def\Lar{\Leftarrow}
\def\bar{\leftrightarrow}
\def\Bar{\Leftrightarrow}

\def\pr{\partial}

\def\Bstar{\bL $\star$ \eL}
\def\etath{\eta_{th}}
\def\irrev{{\mathcal R}}
\def\e{{\rm e}}
\def\noise{n}
\def\hatp{\hat{p}}
\def\hatq{\hat{q}}
\def\hatU{\hat{U}}

\def\iset{\mathcal{I}}
\def\fset{\mathcal{F}}
\def\pr{\partial}
\def\traj{\ell}
\def\eps{\epsilon}




\section{Introduction}
It has been established that one-dimensional
disordered system (1DDS) has a pure point energy spectrum and
its eigenfunctions are exponentially localized in an infinite
system \cite{ishii73,kotani82}. As a result, the ensemble averaged transmission
coefficient of a large enough system decreases exponentially with
respect to the system size $N$ \cite{erdos81}. This statement is established
for standard 1DDS without particular reference to electronic or phonon system
\cite{ishii73}.
On the other hand, it has been  reported that 
in binary disordered systems  special delocalized states exist.
For example, the diagonally disordered dimer
model corresponding to a one-dimensional tight-binding binary
alloy should have extended states the number of which is proportional
to $\sqrt{N}$ for a finite system size \cite{phillips91}. 
In addition to this, a set of extended modes close to the critical
frequency has been confirmed in the disordered one-dimensional  dimer
harmonic chain \cite{adame93,sanchez94}.

It is difficult to experimentally investigate 
the effect of structural correlation on the localization of 
one-dimensional the electronic systems
due to the electron-electron interaction.
The correlation effect have been experimentally 
realized by using a one-dimensional 
optical systems \cite{barthelemy08,marin14,choi17}.
Indeed, in recent years, there have been experimental studies on 
photonic localization in disordered glasses in which light waves 
perform a L\'{e}vy flight \cite{barthelemy08}
and anomalous localization in microwave waveguide with 
long-range correlated disorder \cite{marin14}.
In our earlier papers, we have also numerically investigated the localization
 phenomena of electronic systems with long-range correlations generated 
by the modified Bernoulli map (MB) with stationary-nonstationary 
chaotic transition (SNCT) \cite{yamada18}. 
The detailed property of the MB map is given in
Refs. \cite{aizawa84,aizawa89,tanaka95,akimoto05,akimoto07,akimoto08}.
In particular, wave packet dynamics in the nonstationary potential  
has been investigated \cite{moura98,yamada15}.
The sequence exhibits asymptotic non-stationary chaos 
characterized by the power spectrum $S(f) \sim 1/f^\alpha$ 
($f<<1$), where $f$ denotes frequency
 and $\alpha$ is spectrum index, for $\alpha >1$.
We shall refer to such a system MB chain 
in the following \cite{aizawa84,akimoto05}.
In the MB chain, it is possible to create the potential sequence that
changes  its property from short-range correlation (SRC) 
including $\delta-$correlations to the long-range correlation (LRC), 
if we regulate the correlation parameter $B$.
The relation between the spectrum index $\alpha$ and the correlation parameter
$B$ of the MB map is given by Eq.(\ref{eq:alpha-B}) in the text.
However, the studies on the properties of the phonon system 
with LRC, as compared with those in the electron system, are rather scare
\cite{lu88,tong99,datta95,fabian97,allen98}. 
In particular, there are many unclear points as to phonon dynamics
of the disordered harmonic chains with LRC 
\cite{shahbazi05,naumis06,bahraminasab07,esmailpour08,richoux09,
garcia10,lepri10,costa11b,sales12,juniora15,albuquerque15,zakeri15}.

In this paper, disordered
phonon systems with  LRC generated by MB map 
are studied numerically.
We aim at reporting the characteristic $B-$dependences of 
the Lyapunov exponent  (L-exponent), and the phonon dynamics.
Although the L-exponent is positive 
throughout the entire set of $B$ regions studied here,  
 the $B-$dependence decreases linearly for $ B <2 $, and it 
decays exponentially for $B> 2$.
The distribution of the L-exponent
over the ensemble for the disorder configuration  exhibits some
 slow convergence as $N \to \infty$, unlike that of uncorrelated disordered
systems because of its LRC.
In addition, we investigate the time-dependence of the initially localized wave packet 
in the MB chain due to the displacement excitation by changing the correlation
parameter.
We confirm that the spread of the energy wave packet exhibits 
subdiffusive behaviour, compared to ballistic one as $B$ increases.



This paper is organized as follows.
In the next section, we shall briefly introduce the phonon model 
and the modified Bernoulli map.
In Sect.\ref{sec:lyap} we report about the 
behaviour of the $B-$dependence 
of  Lyapunov exponent  at some frequencies by the numerical calculation.
We show the correlation of the mass sequence effects on the convergence 
property of the distribution as the system size increases.
In Sect.\ref{sec:dynamics}, phonon dynamics in the MB chains
is investigated by changing the correlation parameter.
The summary and discussion  are presented in the last section.
Appendix \ref{app:transmission} shows
 that the anomalous distribution of 
phonon transmission coefficient over ensemble
have a non-universal form, which is different from that in 
uncorrelated disordered chains.

\section{Model}
\label{sec:model}
Here we consider the harmonic chain model represented by the following equation
of motion:
\beq
m_n\frac{d^2u_n}{dt^2}=-K_{n-1,n}(u_n-u_{n-1}) +K_{n,n+1}(u_{n+1}-u_{n}),
\label{eq:equation-of-motion} 
\eeq
where $u_n$ is the displacement from its equilibrium position of
the $n-$the atom and $m_n$'s and $K_{n,n+1}$'s are sequences of masses and
force constants of nearest neighbour atoms, respectively.  
We deal with two types of phonon model, the one is with disordered masses
but constant force constants $K_{n,n+1}=K(=1)$ (mass model), and the other is
spring disordered model with constant mass $m_n=m(=1)$  (spring model). 
The one model is
transformed into the other model by a dual transformation \cite{toda66}.
In the long wavelength approximation of the mass model,
we obtain a scalar wave equation, using continuous variables $x$, 
\beq
\frac{\partial^2 u(x,t)}{\partial t^2}=\frac{K}{m(x)}\frac{\partial u(x,t)^2}{\partial x^2},
\eeq
and for
the continuous variable version of the spring model we get the following wave equation:
\beq
\frac{\partial^2 u(x,t)}{\partial t^2}=\frac{\partial}{\partial x} 
\left[ \frac{e(x)}{m_0} \frac{\partial u(x,t)}{\partial x} \right],
\eeq
where $e(x)$ is the $x-$dependent elastic stiffness and $m_0$
 is the mass density of the medium.

These phonon systems start to be equivalent to that of off-diagonal 
tightly binding electronic system 
$-t_{n,n+1}v_{n+1}-t_{n-1,n}v_{n-1}+\beta v_{n}=Ev_n$
with constant diagonal
element $\beta$ by a transformation using mass-reduction as follows:
\beq
\begin{cases}
\sqrt{m_n}u_n \to v_n,   \\
-\frac{K_{n,n+1}}{\sqrt{m_nm_{n+1}}} \to t_{n,n+1} .
  \end{cases}
\eeq
Furthermore, the particle-hole symmetry for $E=0$ in the off-diagonal electronic 
system corresponds to the translation mode of $\omega=0$ 
in the phonon system.
For weak correlations, all the eigenmodes with $\omega>0$ are localized.
The uniform mode ($\omega=0$) remains extended in the thermodynamic limit.

In this paper we deal with mass model that 
the mass sequences $m_n$'s is generated by a
modified Bernoulli (MB) map with LRC.  The MB map is
one-dimensional map proposed in order to reveal the statistical
properties of an intermittent chaos \cite{aizawa84},
\beq
 X_{n+1} =  
\begin{cases}
 X_{n} + 2^{B-1}X_{n}^{B}  & (0 \le X_{n} < 1/2)  \\
   X_{n} - 2^{B-1}(1-X_{n})^{B} & (1/2 \le X_{n} \le 1),    
  \end{cases}
\label{eq:map}
\eeq
\noindent
where $B$ is a non-negative bifurcation parameter which controls
the strength of correlation among the sequence $X_n$'s. 
We use the symbolized sequence according to the following rule:
\beq
m_n =
 \begin{cases}
m_a &  0\le X_{n}< 1/2 \\
m_b &  1/2\le X_{n} \le 1.
 \end{cases}
\label{eq:binary}
\eeq
\noindent
as a sequence of the masses. 
The mass ratio $R=m_b/m_a$ stands for the parameter controlling the
strength of the disorder. 
 The power spectrum in the low frequency limit  ( $f<<1$)
and thermodynamic limit ($N \to \infty$)    
$S(f)=\frac{1}{N}\left|\sum_{n=0}^N m_ne^{-i2\pi f n/N}\right|^2$ ($f=0,1,2,...,N-1$) 
should behave as
\beq
S(f) \sim
 \begin{cases}
 f^0  & 1 \le B < 3/2  \\
   f^{-\alpha}  & 3/2 \le B \le \infty, 
 \end{cases}
\label{eq:power}
\eeq
where 
\beq
 \alpha \simeq \frac{2B-3}{B-1}.
\label{eq:alpha-B}
\eeq
The bifurcation parameter $B=1$ exhibits an exponential damping of
the correlation. In the range of $1<B<3/2$, the resulting white power
spectrum proves that the sequence has only SRC. 
The theoretical interpretation of the power spectral density has been
given by renewal process analysis for the semi-markovian symbolic 
dynamics \cite{aizawa84c}. And it has been reported that the numerical result
also supports the theoretical result \cite{aizawa84}.

Here,  it is important to mention the difference between the MB chain 
and the model with L\'{e}vy-type disorder in Ref.\cite{zakeri15}.
There is a common point in that the inverse power-law distributions 
 are used in order to characterize the correlation, but there is a significant difference 
in the nonstationary characterized by the power spectrum $S(f) \sim f^{-\alpha}$
with $\alpha \geq 1$ \cite{zakeri15}.
The L\'{e}vy-type disorder model generates the nonstationary sequence with $\alpha=1$
 at a point of the control parameter of the correlation (L\'{e} exponent), while MB chain
with $B \geq 2$ can generate various  nonstationary sequence with $1 \leq \alpha < 2$.

\begin{figure}[htbp]
\begin{center}
\includegraphics[width=7.5cm]{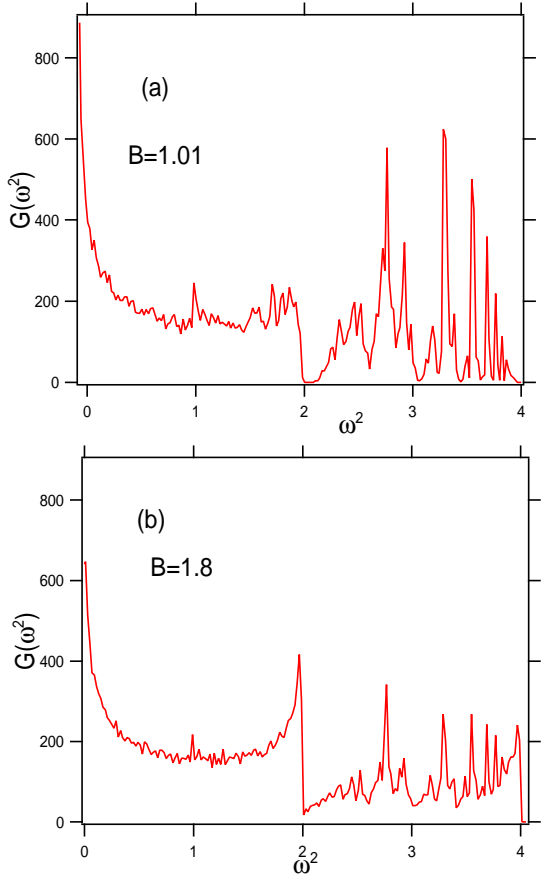}
\caption{
(Color online)
Spectra of squared frequencies for the MB chains 
with $m_a=1, m_b=2$ 
of the bifurcation parameter (a)$B=1.01,$ and (b)$B=1.8$.
We used a fixed boundary condition and the system size is $N=2^{15}$.
The mesh of a horizontal line is $0.02$.
}
\label{fig:phonon-dos-1}
\end{center}
\end{figure}

Assuming the monochromatic time dependence $u_n(t)=e^{-i\omega t}u_n(t=0)$
for Eq.(\ref{eq:equation-of-motion})
 we obtain the stationary equation of motion, 
\beq
-m_n\omega^2 u_n=-K_{n-1,n}(u_n-u_{n-1}) +K_{n,n+1}(u_{n+1}-u_{n}), 
\label{eq:equation-of-motion-2} 
\eeq
characterized by a frequency $\omega$.
The phonon spectrum $G(\omega^2)$ of the mass model ($m_a=1.0,m_b=2.0$) 
is shown in Fig.\ref{fig:phonon-dos-1}. This spectrum for the case of
$B=1.01$ is strongly resembling that in the uncorrelated chains.  
This spectra is characterized by a singular-peak
structure and infinitesimally small gaps dubbed the special
frequencies \cite{hori72,dean72}. A complicated structure in phonon amplitude
and phase is related to this singular structure of the phonon
spectra. We can also observe a structure like the van-Hove
singularity in a periodic lattice in the case of $B=1.8$ as a result of LRC.
 Naturally the same features are observed.
 The details will be reported in elsewhere. 
We keep our eyes on the localization properties
of phonon amplitude in the next section.

\section{Localization of phonon amplitude}
\label{sec:lyap}
In this section, we study the Lyapunov exponent depending on the 
correlation parameter $B$ by using the transfer matrix method
\cite{scales97,yamada01}.

\subsection{Transfer matrix and Lyapunov exponent}
 The equation (\ref{eq:equation-of-motion-2}) can be written 
in terms of the product of the transfer matrix $T_n$ as
\begin{eqnarray}
\left(
\begin{array}{c}
u_{n+1} \\
u_{n} \\
\end{array}
\right)
= T_n
\left(
\begin{array}{c}
u_n \\
u_{n-1} \\
\end{array}
\right)
=\Pi_{i=1}^{n} T_i
\left(
\begin{array}{c}
u_1 \\
u_{0} \\
\end{array}
\right),
\label{eq:transfer-matrix-1}
\end{eqnarray}
where
\begin{eqnarray}
T_n=
\left(
\begin{array}{cc}
\frac{(K_{n-1,n}+K_{n,n+1}-m_n\omega^2)}{K_{n,n+1}}  & -\frac{K_{n-1,n}}{K_{n,n+1}}\\
1 & 0 
\end{array}
\right).
\label{eq:transfer-matrix-2}
\end{eqnarray}
 We are interested in the asymptotic property of the amplitude
$u_n$ for $n \to \infty$ or the corresponding limit theorem
for the product of the matrices.
The asymptotic behaviour of 
Eq. (\ref{eq:transfer-matrix-1}) with respect to the system size 
is characterized by the L-exponent of the 
phonon amplitude of finite size $N$ as follows:
\beq
\gamma_N=\frac{\ln || M(N) \bm{u}_0||}{2N},
\eeq
where $M(n)=\Pi_{i=1}^n T_i$, $\bm{u}_0=(u_1,u_0)^T=(1,0)^T$
when the set of the initial values $u_0$ and $u_1$ is given.  
The frequency dependence of 
the localization length $\xi_N \simeq 1/\gamma_N$ for acoustic 
and electromagnetic waves in a one-dimensional randomly layered media
 is also studied analytically \cite{scales97}.

\begin{figure}[htbp]
\begin{center}
\includegraphics[width=7.0cm]{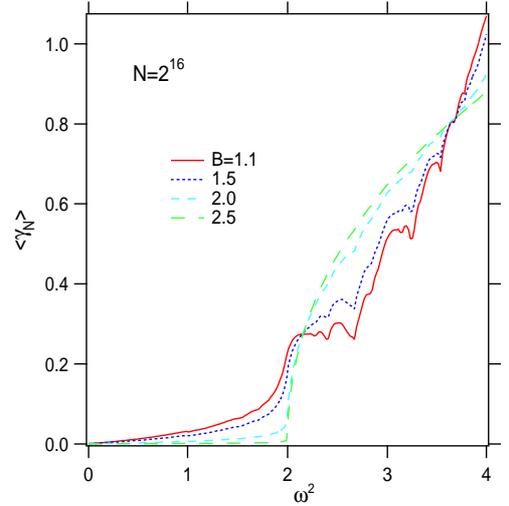}
\caption{
(Color online)
The ensemble averaged 
Lyapunov exponents $\left< \gamma_N \right>$ of MB chains of the size $N=2^{16}$
as a function of  squared frequency at $B=1.1, 1.5, 2.0, 2.5$.
The ensemble size is $2^{11}$.
}
\label{fig:lyap-energy-1}
\end{center}
\end{figure}

Figure \ref{fig:lyap-energy-1} shows the $\omega^2-$dependence of  the 
averaged L-exponents $\left< \gamma_N \right>$ in the MB chains.
The zero mode $\omega=0$ corresponds to the extended state 
in the translational motion mode.
In the case of SRC ($B=1.1$),  it can be confirmed that 
$\gamma_N \propto \omega^2$($\omega <<1$, $N>>1$).
Looking at the smaller side of $\omega$ that works for dynamics, we see 
that L-exponent decreases as $ B $ increases.
The $\omega-$dependence of the L-exponent has been investigated
for the anomalous localization in one-dimensional chains with L\'{e}vy-type 
disorder and the power-law dependence for the L\'{e}vy exponent 
is found \cite{zakeri15}.

\begin{figure}[htbp]
\begin{center}
\includegraphics[width=8.2cm]{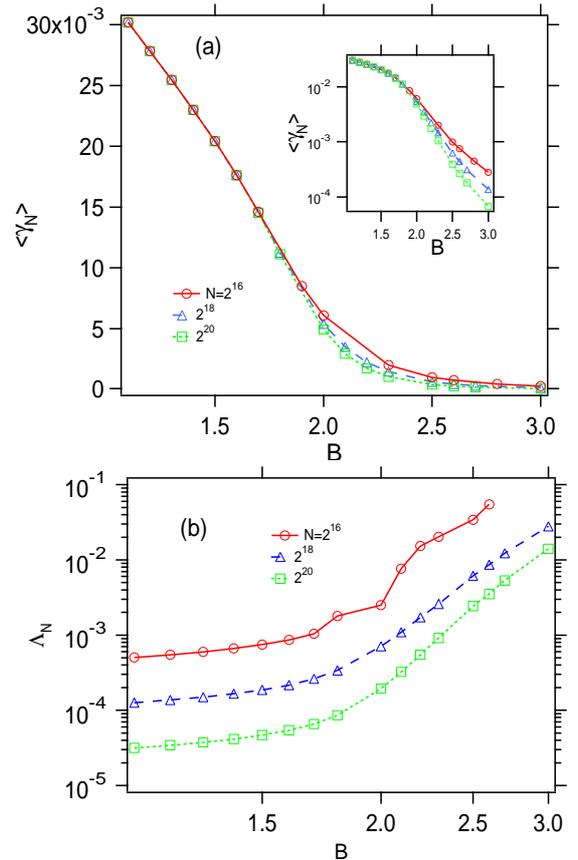}
\caption{
(Color online)
(a)The ensemble averaged 
Lyapunov exponents $\left< \gamma_N \right>$ of the phonon amplitude 
for $\omega^2=1.01$ as a function of the bifurcation parameter.
The cases of $N=2^{16}, 2^{18}, 2^{20}$ are plotted.
The ensemble size is $2^{15}$.
The inset shows the logarithmic plot.
(b)The normalized localization length of the phonon amplitude 
for $\omega^2=1.01$ as a function of the bifurcation parameter. 
}
\label{fig:llyap-av}
\end{center}
\end{figure}

We investigate whether the state is delocalized by increasing $B$ or not.
The $B-$dependence of 
the  averaged Lyapunov exponents $\left<\gamma_N \right>$ 
are shown in Fig.\ref{fig:llyap-av}(a).
It almost linearly decreases  until $B \simeq2$ and 
it experiences exponential decrease up to 
zero $\left<\gamma_N \right>\simeq 0$ for $B>2$.
We examined the case of $\omega^2=1.01$, but similar results are obtained 
for the other frequency in $0 <\omega^2 <2$.
Such a behaviour is resembling that reported for the acustic system 
 in Ref.\cite{costa11a}.
Furthermore, 
the behaviour of the phonon
amplitude is qualitatively similar to that in an electronic system \cite{yamada18}.

%
%
The numerical result of the well-examined electronic systems generated 
by the MB map  strongly suggests that the L-exponent is 
positive even for $B \geq 2$, at least in the investigated range ($2 \leq B \leq 3$).
This result is consistent with many claims that there is no transition
 to delocalization in the regime $\alpha <2$, 
in many electronic systems which are well studied numerically
by using the finite-size scaling 
\cite{moura98,shima04,kaya07,gong10}.
Correspondingly, in the phonon system examined in this paper,  
for  $B \geq 2$ and increasing $N$, the averaged L-exponent does not seem to 
become zero  except for $\omega=0$.
However, as shown in in the Fig.\ref{fig:llyap-av}(a), 
it is suggested that the dependence on $B$ 
changes with $B \simeq 2$ from linear decay to exponential decay,
i.e. $\left< \gamma_N \right> \sim e^{-c(B-2)}$.
This change of the $B-$dependence corresponds to 
the chaos-chaos transition at $B=2$ of the MB map. 

We define the normalized localization length (NLL) to characterize the tail 
of the wavefunction \cite{yamada15,costa11a,shima04,yamada04},  
\beq
\Lambda_N \equiv \frac{1}{\left< \gamma_N \right>N}.
\eeq
It is useful to study the localization and delocalization property 
that $\Lambda_N$ decreases (increases) with the system size 
$N$ for localized (extended) states,
and it becomes constant for the critical states.
The $B-$dependence of 
the NLL are shown in Fig.\ref{fig:llyap-av}(b).
It is found that the even for $2<B (<3)$ the NLL decreases and the localization length 
$1/\left< \gamma_N \right>$ is less than the system size $N$
in the thermodynamic limit $N \to \infty$.
As a result, we can say that the states are exponentially localized for the case of $\omega^2=1.01$. 

lt is worth noting that Furstenberg's theorem can be applied to the
product of matrices for at least the stationary regime $B<2$, 
because the sequence is a renewal process
with a finite average residence time in the MB chain \cite{ishii73}. 
As a result, L-exponent in the infinite system is positive-definite and sample
independent with probability l for any non-vanishing and finite
initial vector \cite{goda91,yamada96}. 

More detailed investigation is necessary
in the nonstationary regime with the power spectrum $S(f) \sim f^{-\alpha}$ 
with $1\leq \alpha<2$ because the nonstationary sequence
 is not the sufficient condition for the delocalization.

\subsection{Convergence of the distribution}
Previous section demonstrates, 
the $B-$dependence of the average value of L-exponent is similar to 
that of the electronic system.
The correlation effect is also expected to show up in the form at the 
anomalous distribution and variance,  as in the case of electronic system.
Therefore, we will confirm the point.


\begin{figure}[htbp]
\begin{center}
\includegraphics[width=9.0cm]{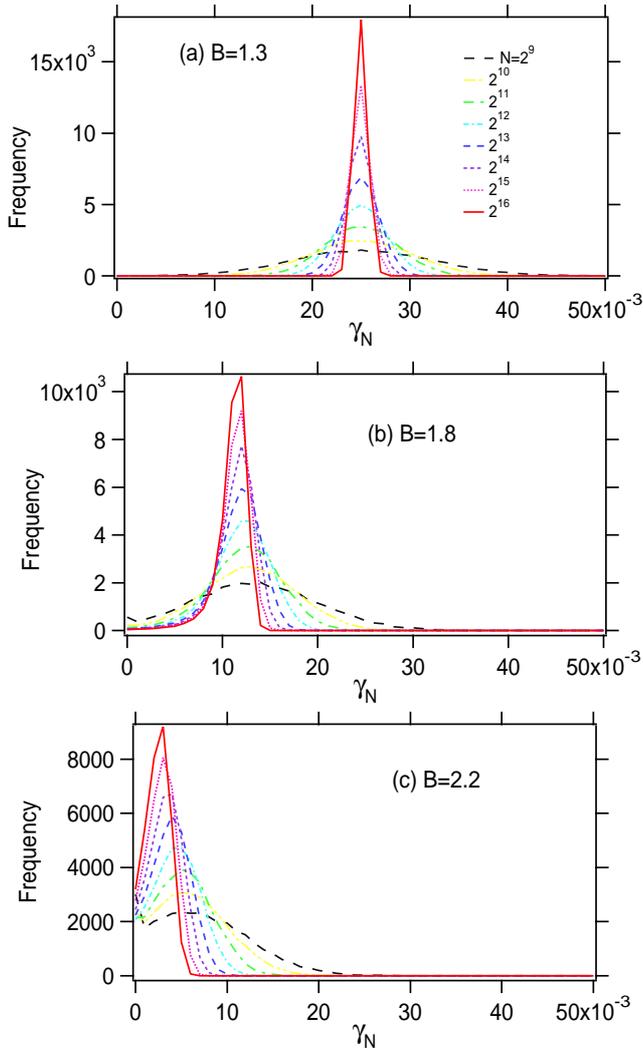}
\caption{
(Color online)
Histograms of the distribution of L-exponent of phonon
amplitude in the MB chains with some system size $N=2^9-2^{16}$
for squared frequency
$\omega^2=1.01$ in the cases of   (a)$B=1.3$, (b)$B=1.8$ and $B=2.2$.
 We have used mass model with $ma=1, m_b=2$
 and a mesh of histogram in a horizontal line is $0.001$.
The ensemble size is $2^{15}$.
}
\label{fig:yaphis-101}
\end{center}
\end{figure}

\begin{figure}[htbp]
\begin{center}
\includegraphics[width=9.0cm]{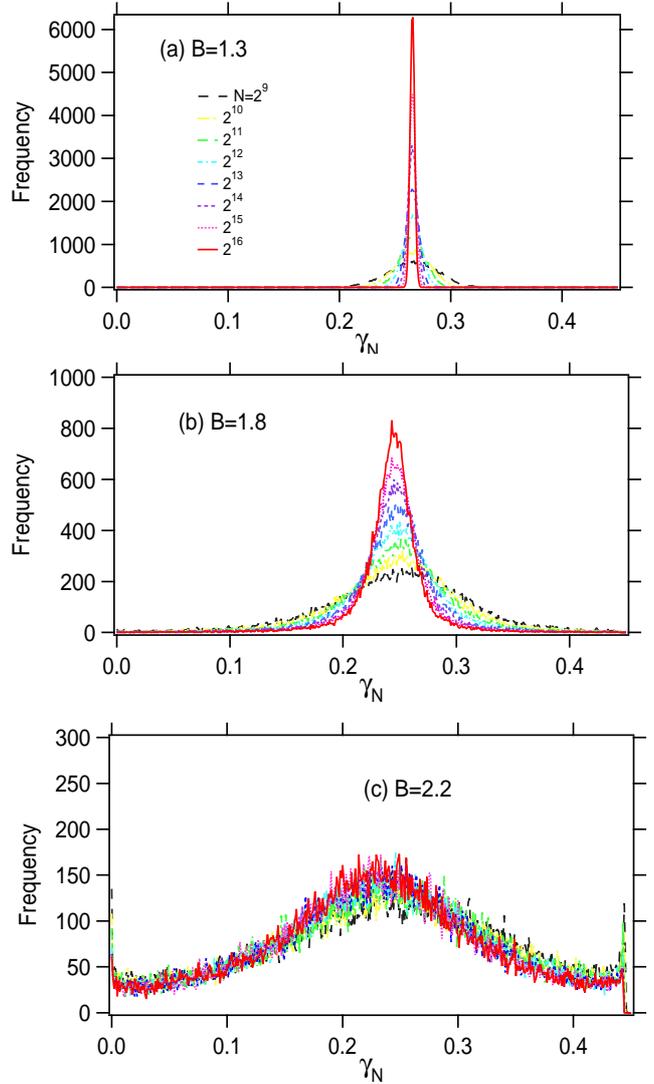}
\caption{
(Color online)
Histograms of the distribution of L-exponent of phonon
amplitude in the MB chains with some system size $N=2^9-2^{16}$
for squared frequency
$\omega^2=2.1$ in the cases of   (a)$B=1.3$, (b)$B=1.8$ and $B=2.2$.
 We have used mass model with $ma=1,m_b=2$
 and a mesh of histogram in a horizontal line is $0.001$.
The ensemble size is $2^{15}$.
}
\label{fig:yaphis-21}
\end{center}
\end{figure}


  Figures \ref{fig:yaphis-101} and  \ref{fig:yaphis-21} 
show the distribution of L-exponent of phonon
amplitude over $2^{15}$ samples. 
We have performed the calculations for 
the mass model with a mass ratio $(R=2)$ at some values of
squared frequency $\omega^2$.  
Distribution of almost Gaussian type are
observed in the case of $B=1.3$
, in which the structural
correlation is of short range. The behaviour of the distribution
is quite similar to that of the uncorrelated disordered system, while
the distribution obeys normal central-limit theorem (CLT). 
The multi-peak structure is observed in the distributions for the case
of $B=2.2$ at the squared frequency $\omega^2=2.1$.  
Two sharp peaks on the both sides of 
the main distribution in Fig.\ref{fig:yaphis-21}(c) 
 come from the LRC, which proves a certain amount of very long pure and
almost pure subsystems. 

\begin{figure}[htbp]
\begin{center}
\includegraphics[width=7.5cm]{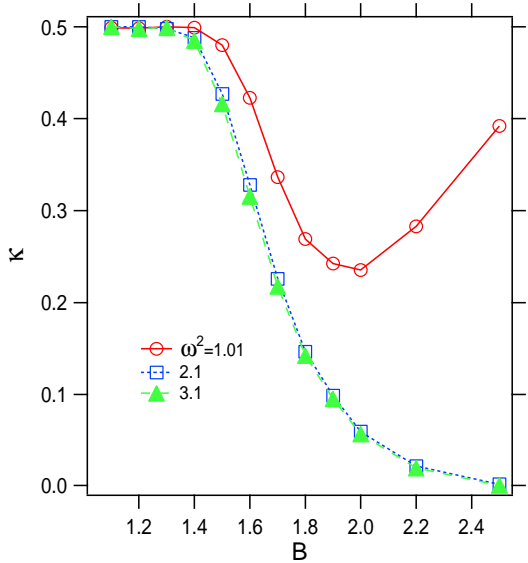}
\caption{
(Color online)
The exponent $\kappa(B)$ of the power for the standard deviation of
the L-exponent of the phonon amplitude as a function of the
bifurcation parameter.
The cases of $\omega^2=1.01,2.1,3.1$ are plotted.
}
\label{fig:yap-sd-1}
\end{center}
\end{figure}

  We consider the fluctuation of L-exponent distribution using
the scaling form, 
\beq
\sqrt{\left<(\Delta \gamma_N)^2\right>} \propto N^{-\kappa(B)}, 
\eeq
to fit the numerical data.
The estimated value of  $\kappa(B)$ is plotted in Fig.\ref{fig:yap-sd-1}. 
For $1<B<3/2$,
the value of  $\kappa$ is roughly $1/2$, implying that the convergent
property of the distribution with respect to $N$ obeys or
approximately obeys CLT. However, for $3/2<B<2$, the distribution
converges more slowly than that obeying the CLT. This property
is a remarkable feature as the correlation increases.
Moreover, convergence of distribution is hardly observed for $B \geq 2$.
In particular, in the case of $\omega^2=2.01$, it accumulates to 
$\left< \gamma_N \right> \simeq 0$, so variance of distribution increases.
As a result, the $B-$dependence of the L-exponent and 
the convergence property of distribution form change 
around SNCT $B \simeq 2$ of the MB map.

The apparently anomalous distribution of the L-exponent is also reflected in 
the distribution of the transmission 
coefficient which is closer to the physical quantity. 
The data showing the abnormality of the distribution of the phonon transmission coefficient 
different from the usual uncorrelated disordered ones are in the Appendix.

\section{Phonon dynamics}
\label{sec:dynamics}
In this section, we study the time-evolution of vibrational wave packet 
described by the equation of motion in 
Eq.(\ref{eq:equation-of-motion})
on the MB chains with the long-range correlated 
random masses by using the leap-frog integration scheme \cite{okabe96}
with the time mesh $\Delta t=0.0001$.

As initial conditions ($t=0$), we chose a delta function at middle site of the chain, 
and the zero velocities at all the sites as follows;
\beq
 \begin{cases}
u_n(t=0)=U_0\delta_{n,N/2} \\
v_n(t=0)=0.
 \end{cases}
\label{eq:initial}
\eeq 
Further we chose $U_0=1$.
Although other choices are possible, 
the qualitative conclusions of this report are independent of
the initial conditions. 
To qualitatively measure the degree of localization of the phonon, 
we evaluate the spreading of the wave packet by 
the mean square displacement of the phonon amplitude $m_2(t)$ \cite{naumis06}, 
the participation ratio of the amplitude $P(t)$ \cite{lu88,tong99},
and the energy spread of the wave packet 
$E(t)$ \cite{datta95,allen98,albuquerque15} 
as a function of time.
Below, we see how these three quantities change when $B$ increases, 
compared with a case of binary periodic chain (BP).

\subsection{Mean square displacement  $m_2(t)$}
First, we compute the time-dependence of the second order moment;
\beq
m_2(t) = \frac{1}{N} \sum_{n=1}^N \left(n-\frac{N}{2} \right)^2 \left< |u_n(t)|^2 \right>, 
\label{eq:msd}
\eeq
where $\left<..\right>$ denotes the average over the initial values of the MB map.

\begin{figure}[htbp]
\begin{center}
\includegraphics[width=8.0cm]{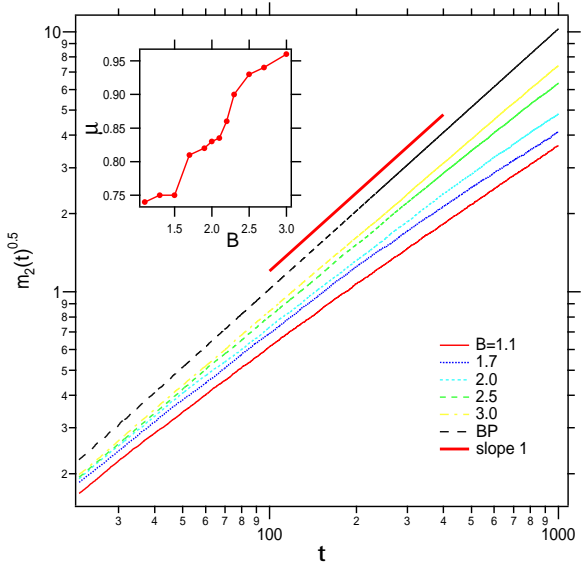}
\caption{
(Color online)
The time evolution of the wave packet  spreading  $\sqrt{m_2(t)}$
of the phonon amplitude in the MB chains with the initial condition
(\ref{eq:initial}) for $B=1.1,1.7,3.0$.
The inset shows the index $\mu$ as a function of $B$, 
estimated by numerical data of  $\sqrt{m_2(t)}$.
Result of the binary periodic chain (BP) is also plotted as a reference.
Note that the data are plotted in double logarithmic scales.
The ensemble size is $100$.
}
\label{fig:m2(t)-1}
\end{center}
\end{figure}

Figure \ref{fig:m2(t)-1} shows the numerical result  of the root mean square 
displacement $\sqrt{m_2(t)}$ in the MB chains with $B=1.1,1.7,3.0$. 
The log-log plot of the data reveals $\sqrt{m_2(t)} \sim t^\mu$, and 
numerically estimated $\mu$ is shown in the inset.
It occurs that for weak correlation cases it is subdiffusive $0.7 \lesssim \mu<1$,
and the index $\mu$ gradually approaches 1 as the $B$ increases,
which proves the ballistic spreading.
This tendency agrees with that reported by Naumis {\it et al} \cite{naumis06}.
Accordingly, to sum up, it can be said that the difference due to 
the change in B is small.

\subsection{Participation ratio $P(t)$}
We define the time-dependent participation ratio (PR) of the displacement
$u_n(t)$ by the following equation,
\beq
P(t) &=& \frac{\sum_{n=1}^N \left<|u_n(t)|^2 \right>}{\sum_{n=1}^N \left< |u_n(t)|^4 \right>}, 
\label{eq:pr}
\eeq
It is clear that $P(t) \simeq \xi_{p}^2$ for an exponentially localized case, 
where $\xi_{p}$ is the localization length of the wave packet.
On the other hand, if the wave packet is extended $P(t)$ will be of order of unity,
$P(t) \sim O(1)$ \cite{lu88}.

\begin{figure}[htbp]
\begin{center}
\includegraphics[width=7.5cm]{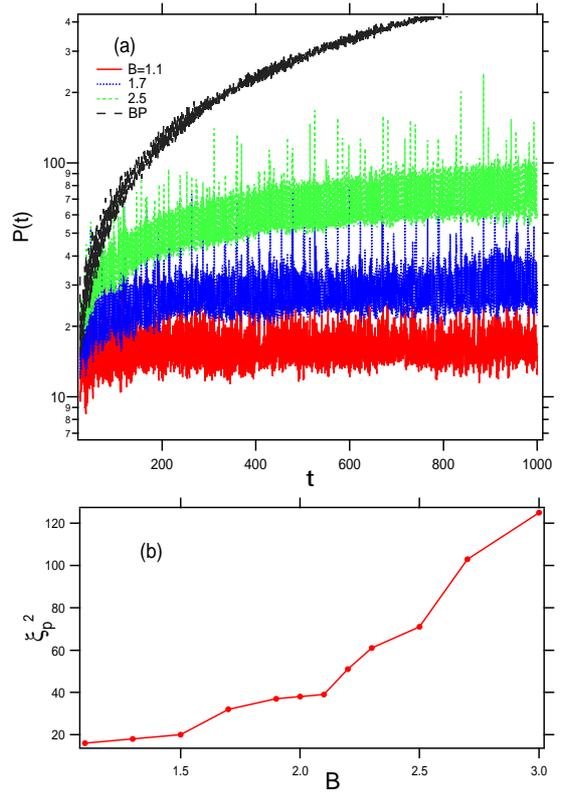}
\caption{
(Color online)
(a)The time-dependence of  the participation ratio $P(t)$
in the MB chains with the initial condition
(\ref{eq:initial}) for $B=1.1,1.7,3.0$.
Result of the binary periodic chain (BP) is also plotted as a reference.
Note that the data are plotted in logarithmic scales.
The ensemble size is $100$.
(b)The squared localization length $\xi_p^2$ of the phonon amplitude
which is numerically estimated by the participation ratio for $t>>1$.
}
\label{fig:P(t)-1}
\end{center}
\end{figure}

Figure \ref{fig:P(t)-1}(a) shows time-dependence of the PR in the logarithmic scale.
It follows that the localization length $\xi_p$ increases as $B$ increases, and 
it is very different from the behavior of the binary periodic system.
In Fig.\ref{fig:P(t)-1}(b), the squared localization length $\xi_p^2$ of 
the phonon amplitude is shown,  
which is numerically estimated by $P(t) \simeq \xi_{p}^2$ for $t>>1$.
It can be seen that $\xi_{p}^2$ increases gradually for $B<2$ 
and rapidly increases for $B>2$ as $B$ increases.

\subsection{Energy diffusion  $E(t)$}
The energy of the disordered chain is distributed in a time-varying fashion between 
their kinetic and potential energies.
Here, we can define the actual mid-value of the energy of the pulse as follows: 
\beq
R(t)=\frac{\sum_{n=1}^N n \left<E_n(t)\right>}{\sum_{n=1}^N\left<E_n(t)\right>}, 
\eeq
where $E_n(t)$ denotes the local energy  at site $n$.
Furthermore, the spread of the energy for the initial displacement excitation
can be defined by the second order moment as follows;
\beq
E(t) &=& \frac{\sum_{n=1}^N \left(n-R(t) \right)^2\left<E_n(t)\right>}{\sum_{n=1}^N\left<E_n(t)\right>}.
\label{eq:e-diff}
\eeq
Although the fluctuation of the position $R(t)$ of the centre of the energy exists, 
the energy spreading Eq.(\ref{eq:e-diff}) is essentially the same as 
in the case of $\nu=2$ in the Eq.(47) in the reference \cite{lepri10},
It  has been reported that in the uncorrelated random chains
while the participation number remains finite, i.e. localized state, 
the energy spread is shown to be way 
$E(t) \sim t^{0.5}$ (short-wavelength limit) after displacement excitation
owing to the unscattered states of the order $O(\sqrt{N})$ 
around $\omega=0$ \cite{datta95}.
In periodic chains,   $E(t)$ exhibits the ballistic spread as  $E(t) \sim t^{2}$.

\begin{figure}[htbp]
\begin{center}
\includegraphics[width=8.0cm]{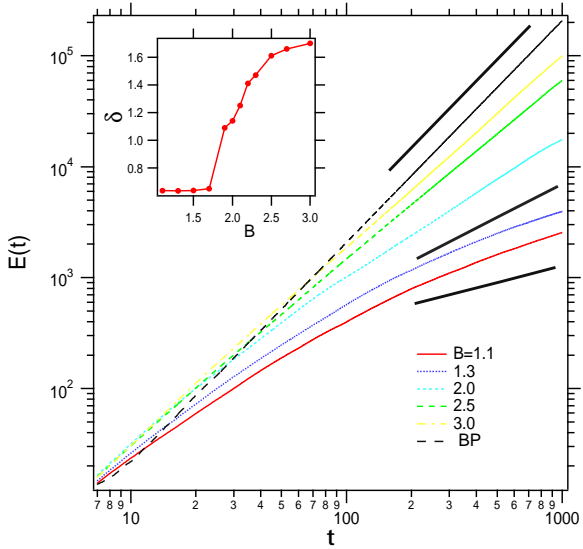}
\caption{
(Color online)
(a)The time-dependence of  the energy spread $E(t)$
in the MB chains with the initial condition
(\ref{eq:initial}) for $B=1.1,1.7,3.0$.
Result of the binary periodic chain (BP) is also plotted as a reference.
Note that the data are plotted in double  logarithmic scales.
The ensanble size is $100$.
Three bold lines correspond to $t^{0.5}$, $t^1$ and $t^2$, respectively.
The inset shows the power law index $\delta$ as a function of 
the bifurcation parameter $B$ 
which is numerically estimated by the time-dependence of the energy spread
$E(t)$. 
}
\label{fig:E(t)-1}
\end{center}
\end{figure}

It is expected, as shown in Fig.\ref{fig:E(t)-1}, that 
the energy spread asymptotically approaches
the behaviour $E(t) \sim t^{0.5}$ appears at infinite time 
for $B=1.1$. This corresponds to the case of uncorrelated 1DDS. 
Numerically,  it is consistent with the result of the references \cite{zakeri15,albuquerque15}.
Furthermore, it seems that the  
transition from $E(t) \sim t^{0.5}$ behavior to $E(t) \sim t^2$ is occurring at $B$ large.
The $B-$dependence of the index $\delta$ evaluated 
by fitting for $E(t) \sim t^\delta$ is shown in the inset in Fig.\ref{fig:E(t)-1}.
It turns out that $\delta$ rapidly increases from $\delta \simeq 0.5$
and gradually increases toward 
 $\delta \simeq 2$  in the case of 
the binary periodic systems, as $B$ becomes large.
At least in the SRC regime ($B<3/2$) it is $\delta \simeq 0.5$, 
but more detailed numerical investigation is needed 
on how to increase $\delta$ for $B> 3/2$.

\section{Summary and discussion}
We have studied here some statistical properties of L-exponent of
phonon amplitude in a one-dimensional disordered harmonic chain 
with LRC, which is generated by MB map.
The Lyapunov exponents are positive-definite except for the zero mode
$\omega=0$.
the $B-$dependence of the L-exponent and 
the convergence property of the distribution clearly change 
around the SNCT $B \simeq 2$ for the MB map.
The convergence properties of the
distribution of those quantities with system size $N$ do not obey
the central-limit theorem at least for $B>3/2$. 
As $B$ increases, the convergence becomes more slow.
The slow
convergence corresponds to the anomalous large deviation property
of the symbolic sequence \cite{aizawa89}.

Moreover, here we have investigated the phonon dynamics of the initially localized 
displacement excitation in the correlated disordered chains.
There is a tendency that the participation ratio 
of the phonon amplitude is maintained at its finite value even if 
the correlation parameter $B$ increases, i.e. localized state persists.
On the other hand, it has been found that 
the spread of the local phonon energy changes from 
the behavior, $t^{0.5},$ to the ballistic one,  $t^{2}$, 
along with the increase of $B$.
The diffusion index $\delta$ rapidly increases 
around SNCT $B \simeq 2$ of the MB map.
Still, for the phonon dynamics  it is necessary to obtain a 
more detailed numerical result.



In this report, we dealt with the harmonic phonon system, but, 
if the anharmonic terms are introduced in the phonon dynamics, 
localization effect due to the existence
of the breathers modes should also be  expected to occur.
Finally, the effect of the long-rang correlation on thermal conduction should also
be an interesting feature to investigate \cite{lepri03,iubini18}.


The localization phenomenon in the disordered system 
with strong correlation appears in various natural phenomena 
regardless of the electron and phonon systems.
For example, seismic wave propagation in a heterogeneous  rock 
can also be localized due to the multiple scattering and 
interference of the wave \cite{sahmi05}.
We expect that the present work would stimulate further studies 
of the localization in the diverse systems.

\appendix

\section{Anomalous distribution of the phonon transmission coefficient}
\label{app:transmission}
In the same way as in the case of electronic system 
\cite{stone81,mello87,stone91,yamada91,nishiguchi93}, 
this appendix should present investigations of  the correlation effect on 
the statistical property of the phonon transmission coefficient (PTC)
of a finite chain.
We consider a finite chain embedded into an infinite perfect lattice 
with a constant mass ($m=1$), as compared with those in the 
uncorrelated cases ($B=1.1$).
It depends only on the transfer matrix $M(N)$ itself and is 
independent of the boundary condition.
The PTC $T(N)$ of a finite system $N$ is given as, 
\beq
 T(N)= 
 \frac{4\sin^2 K}
{
|-M_{11}e^{-iK}+M_{21}-M_{12}+M_{22}e^{iK}|^2, 
}
\eeq
where $M_{i,j}$ denotes the $i,j$ matrix element of the transfer
matrix $M(N)$ in Eq.(\ref{eq:binary}) and the $K$ means wavevector of incident wave
from semi-infinite perfect lattice with the lattice constant.
The dispersion relation in the perfect lattice is $\omega^2=2(1-\cos K)$ \cite{stone81}.

\begin{figure}[htbp]
\begin{center}
\includegraphics[width=7.0cm]{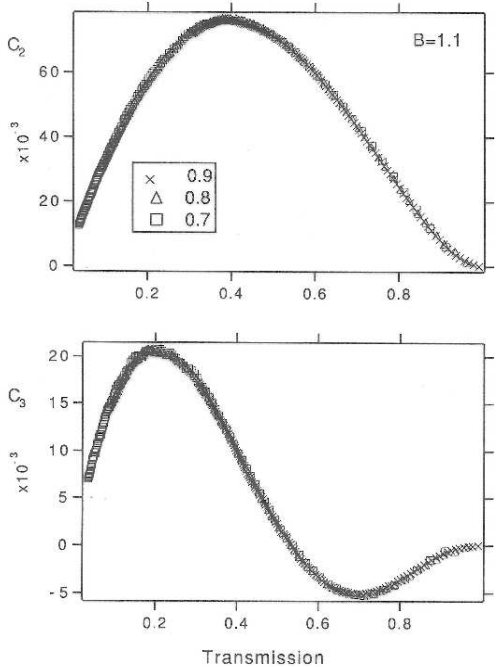}
\caption{
(Color online)
Numerical results of the second- and the third-order cumulant
($C_2$, $C_3$) of the distribution of the PTC at $\omega^2=2$ as a function of
the first-order cumulant for MB chains with the bifurcation
parameter $B=1.1$.
The cases of $m_a=1$ and $m_b=0.9, 0.8$ and $0.7$ are plotted. 
}
\label{fig:Fig4}
\end{center}
\end{figure}

  For the sake of understanding the anomalous feature of the
distribution over ensemble in detail, we study the relation
between the cumulants of distribution for PTC. Some typical
relations between the cumulants for the uncorrelated 
case $B=1.1$ are shown in Fig.\ref{fig:Fig4} \cite{yamada91}. It is well observed that some
data are plotted on universal curve regardless of the mass ratio.
Though these data is only an example of the case of an for a
incident wave with a squared frequency $\omega^2=2$, we can confirm that
it is true also for the other cases when the bifurcation
parameter is in white-power-spectrum regime $(1< B <3/2)$. This
kind of universality has been strongly suggested to exist in electronic
random 1DDS by using several methods \cite{mello87}.

\begin{figure}[htbp]
\begin{center}
\includegraphics[width=7.0cm]{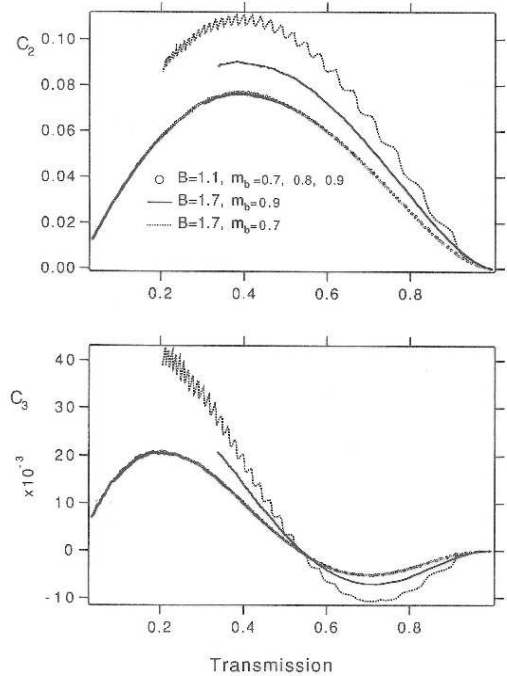}
\caption{
(Color online)
The cumulant relations for MB chains with the bifurcation
parameter $B=1.7$. The result of $m_a$=1 and $m_b=0.7, 0.9$
are plotted.
The data of $B=1.1$ in Fig.\ref{fig:Fig4} are added as a reference.
}
\label{fig:Fig5}
\end{center}
\end{figure}

On the other hand, some remarkable deviations are observed in
the relations between stimulants for some values of the
bifurcation parameter ($3/2< B <2$) in Fig.\ref{fig:Fig5},  compared with the
universal one. Here, we can say that the distribution of the PTC
in MB system does depend on the bifurcation parameter $B$
controlling the structural correlation and it also depends on the
mass ratio $R=m_b/m_a$ in 1DDS with LRC.


Recently, it has been theoretically and experimentally  investigated 
that the anomalous localizations on the transmission of electron 
and the distribution are caused by the correlated disorder 
\cite{falceto10}.

\section*{Acknowledgments}
The author would like to thank Professor M. Goda for discussion 
about the correlation-induced delocalization at 
early stage of this study, 
and  Professor E.B. Starikov for proof reading of the manuscript.
The author also would like to acknowledge the hospitality of 
the Physics Division of The Nippon Dental University at Niigata
for  my stay, where part of this work was completed.

\section*{Author contribution statement}
The sole author had responsibility for all parts of the manuscript.




\end{document}